\documentclass[a4paper]{article}
\usepackage[utf8]{inputenc}
\usepackage[a4paper, total={6in, 8in}]{geometry}
\usepackage[english]{babel}
\usepackage{anyfontsize}
\usepackage{csquotes}
\usepackage{booktabs}
\usepackage{tabularx}
\usepackage{graphicx}
\usepackage{dirtytalk}
\usepackage{hyperref}
\usepackage{float}
\usepackage{multirow}
\usepackage[%
  backend=bibtex      
 ,style=numeric-comp  
 ,sorting=none        
 ,sortcites=true      
 ,block=none
 ,indexing=false
 ,citereset=none
 ,isbn=true
 ,url=true
 ,doi=true            
 ,natbib=true         
]{biblatex}
\newlength\defaultparindent
\AtBeginDocument{\setlength\defaultparindent{\parindent}}
  
\title{Selected Results from the REDMARS2 Project: Recursive Delay-Tolerant Networking using Bundle-in-Bundle Encapsulation}

\author{
    Marius Feldmann,
    Tobias Nöthlich,
    Felix Walter,
    Maximilian Nitsch,\\
    Juan A. Fraire,
    Georg A. Murzik,
    Fiona Fuchs\\
    D3TN GmbH, Germany\\
    \texttt{contact@d3tn.com}
}

\begin{document}

\maketitle

\begin{abstract}
This whitepaper presents parts of the results of the REDMARS2 project\footnote{\url{https://www.forschung-it-sicherheit-kommunikationssysteme.de/projekte/redmars2}} conducted in 2021-2022, exploring the integration of Recursive Internetwork Architecture (RINA) concepts into Delay- and Disruption-Tolerant Networking (DTN) protocols. Using Bundle-in-Bundle Encapsulation (BIBE), we implemented scope-based separation mechanisms resulting in scalable DTNs. A key contribution of this work is the demonstration of practical BIBE-based use cases, including a realistic Solar System Internet communication scenario involving unmanned aerial vehicles (UAVs) and satellite relays. The evaluation, supported by field tests in collaboration with the European Space Agency (ESA), confirmed the viability of BIBE as a foundation for scalable, recursive, and interoperable DTN architectures.
\end{abstract}

\section{Introduction and Overview}
\label{introduction}
In contrast to the largely standardized foundational protocols of the Internet, communication architectures in so-called \say{challenged networks} --- that is, networks characterized by intermittent connectivity and high latency --- are still in the process of standardization.
These networks are found in various application areas, such as underwater networks (see e.g. \cite{coldsun}) and low-earth-orbit satellite networks without a constant connection to ground stations (see e.g. \cite{opssat}).


The REDMARS2 project aimed to investigate a transfer of concepts from the Recursive Internetwork Architecture (RINA) \cite{day2008patterns} approach to the domain of Delay- and Disruption-tolerant Networks (DTN). A core goal was to improve scalability of the overall network while solving issues related to naming and addressing and facilitating the effective use of different routing, forwarding, and discovery mechanisms in separate yet interconnected parts of the network. For this, a prototypical implementation was built that uses the core concepts of RINA within the DTN domain. This prototype was evaluated in a field test conducted with support by the European Space Agency (ESA).

The fundamental idea discussed herein was presented in an interim meeting of the IETF DTN working group in 2022 \cite{rina-video} and an extended overview can be found in a dedicated video \cite{ud3tn-trilogy-redmars-video}.

This whitepaper is divided into five sections, beginning with this Introduction (Section~\ref{introduction}). Section~\ref{concept-bibe} presents the recursive layering approach using BIBE, while its implementation is discussed in Section~\ref{implementation}. The field-test evaluation of the solution is described in Section~\ref{evaluation}. Finally, Section~\ref{summary} summarizes the whitepaper and provides an outlook.

\section{Recursive Networking Stack Based on Bundle-in-Bundle Encapsulation (BIBE)} \label{concept-bibe}
\subsection{Background and Use Case}
The applied network's recursive layering enables the flexible establishment of large-scale overlay networks. These networks (scopes) are stacked on top of each other, bridging lower layers of the network\footnote{Readers may consider the overall concept of subdividing a DTN network related to the concept of \emph{regions}. See Section \ref{concept-regions} for a brief discussion of the relationship between the two.}. A potential use case for this is the Solar System Internet: the previously mentioned overlay network may contain the planets of the solar system, without any specific information about reachability or the network topology of those planets. A scope below the overlay network could then provide information on relay satellites between the planets, allowing for interplanetary communication. At an even lower layer, planet-specific topology information could be available.

A scenario that demonstrates the relevance of this approach is the control of unmanned aerial vehicles (UAVs) on Mars by an operator on Earth. In this scenario, messages transmitted from the operator to the drone are transferred via an overlay network which bridges several underlying transfer networks. After transferring the messages via an underlay network spanning from Earth to Mars, they are forwarded to the drone through the Mars underlay network.

\subsection{Recursive Layering in DTN}
 
For DTNs, the recursive layering can be implemented using Bundle-in-Bundle Encapsulation (BIBE). BIBE is a Bundle Protocol (BP) convergence layer protocol whose services encapsulate an outgoing Bundle into a BIBE protocol data unit for transmission as the payload of another Bundle. \cite{ietf-dtn-bibect-03} This enables modeling the scopes mentioned above using DTN protocol implementations with BIBE Convergence Layer Adapters (BIBE-CLAs). A node on which a number \emph{P} of protocol implementations with BIBE-CLA are running can thus be part of up to \emph{P} scopes. It is possible to pass Bundles on to lower or higher layers by encapsulating and unpacking them. These layers can be completely independent in how Bundles are routed and dispatched.

It shall be noted that for achieving clean and independent layering, one important relaxation of the BIBE specification must be made: it must be possible to direct BIBE Bundles to non-administrative endpoints, such that the BIBE Protocol Data Unit (BPDU) including any processing of the inner Bundle does not have to be handled by the same Bundle Protocol Agent (BPA) --- otherwise, information would necessarily be shared between the layers.
In our implementation, the BIBE-CLA can register like any other BP application with the application interface of the BPA\footnote{Note: This can be the same or (preferably) a different BPA instance than the one that handled the outer (BIBE) Bundle.} and, thus, be completely independent from any Bundle-related processing on the lower layer.

Suppose a layer determines that further information is needed to forward the Bundle to a specific next hop. In that case, the Bundle can simply be passed to a lower layer where more specific routing information is available.

A further advantage of encapsulation is that it prevents routing information from migrating between different scopes. This is because a DTN protocol implementation within a specific scope only processes and evaluates the outermost layer of an encapsulated Bundle. This mechanism ensures a clean separation of the individual scopes.

\begin{figure}[H]
    \centering
    \includegraphics[width=0.8\textwidth]{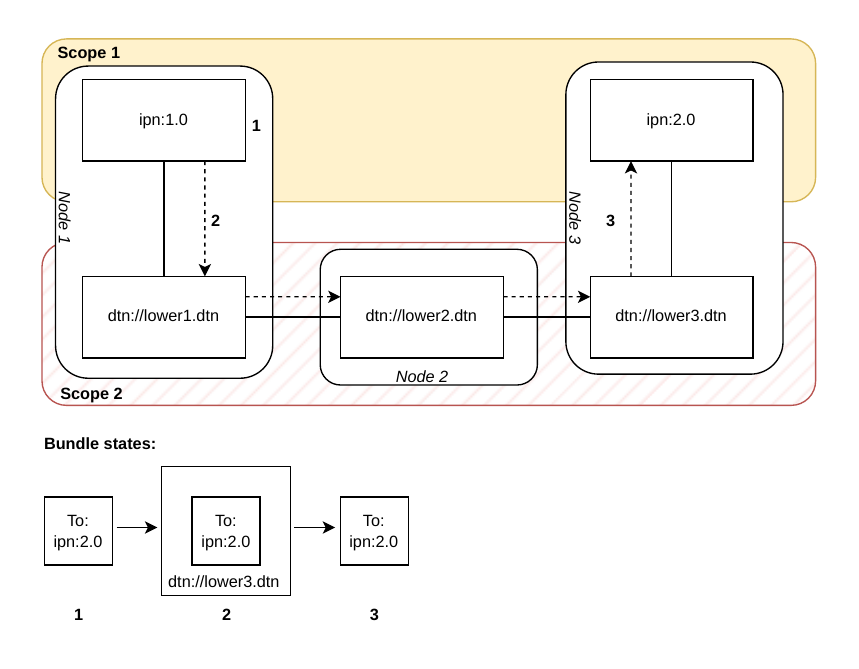}
    \caption{A simple scenario with 3 nodes and 2 scopes. In addition, the encapsulation of the initially created Bundle during transition to a lower scope and the decapsulation during transition to a higher scope are illustrated.}
    \label{fig:bibe_rina_example}
\end{figure}

To make this concept more comprehensible, we provide a simple scenario. The topology of this example is shown in Figure~\ref{fig:bibe_rina_example}. It essentially consists of three nodes\footnote{Note that a \emph{node} in this case does not necessarily mean a single Bundle Node. Indeed, in our implementation, we instantiate two completely separate BPAs that do not possess any configuration information related to the other scope. Only the upper-layer instances' BIBE-CLAs are configured with the application endpoint of the lower-layer BPA plus the lower-layer destination endpoint (``CLA address'') for the next hop.} operating across two scopes, with one node located exclusively within the lower scope. Furthermore, the nodes in the upper scope have no direct connections, so the lower scope must be used to forward a Bundle.

The process of sending a Bundle from Node 1 to Node 3 must therefore begin with the initial creation of a Bundle on Node 1 within Scope 1. Since Node 1 in Scope 1 is unaware of a path to Node 3, the Bundle is passed down to the lower-layer Scope 2.

This transition is accomplished using BIBE, where the initially created Bundle (with the destination EID \texttt{ipn:2.0}) serves as the payload of a subsequent Bundle destined for the EID \texttt{dtn://lower3.dtn}.

It is required that the DTN protocol agent instance of the destination node is addressed in the lower scope in this initial step. This is because Scope 2 cannot process the EID \texttt{ipn:2.0} and, crucially, this EID is never actually evaluated within Scope 2 due to the Bundle's encapsulation. The exact method for determining the destination EID of the encapsulating Bundle is implementation-dependent (like any regular convergence-layer next hop configuration). For this scenario, we assume the EID is statically defined and assigned when the encapsulating Bundle is created.

Once the Bundle enters Scope 2, its processing occurs entirely independently of the encapsulated Bundle. As previously noted, this inner Bundle is never evaluated or parsed within Scope 2. The only point of interaction with the inner Bundle is Node 3, where the subsequent scope transition --- this time into the higher-layer Scope 1 --- takes place. At Node 3, the outer Bundle is removed, and the inner Bundle is then forwarded to the protocol instance associated with the EID \texttt{ipn:2.0}. The various states of the Bundle are summarized in simplified form in Figure~\ref{fig:bibe_rina_example}.

\begin{figure}[h]
    \centering
    \includegraphics[width=\textwidth]{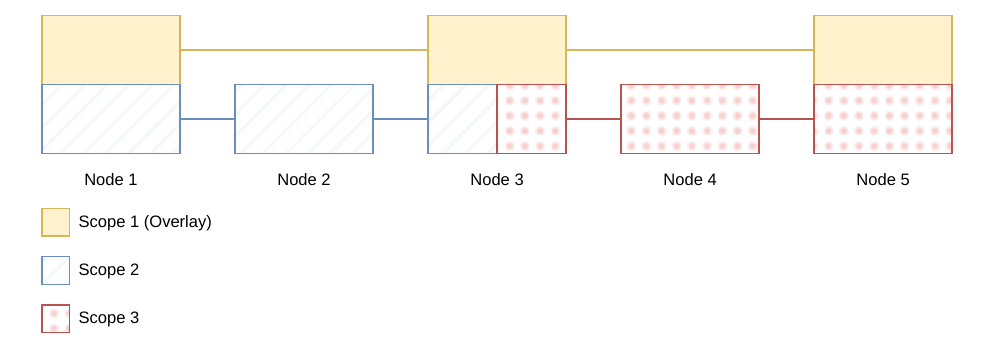}
    \caption{An example scenario with 5 nodes, where there is an overlay scope and two underlying scopes on the lower layer, which do not share any topological or routing information. Accordingly, there are three independent sets of topology information.}
    \label{fig:rina}
\end{figure}

\subsection{Routing} \label{concept-routing}

Since routing is bound to individual scopes, routing information does not propagate across them. In principle, different routing algorithms and approaches can be used in each scope, depending on the network's size and nature. In addition, the destination EIDs of a Bundle differ depending on the scope. This can be easily illustrated with the help of Figure~\ref{fig:rina}. A Bundle that is to be sent from Node 1 to Node 5 must use Node 3 as an intermediate hop. 
Since the DTN overlay scope lacks information on how exactly the next hop can be reached, and Node 3 can only be reached via Node 2, which is not known to the DTN overlay network, the Bundle is transferred to Scope 2 (the lower scope) with the new destination Node 3. Within Scope 2, the Bundle can now be forwarded via Node 2 to Node 3. Since Node 3 is the destination node of the encapsulating Bundle, it is decapsulated and passed to the higher-level Scope 1, where it is determined that Node 5 is the actual destination node. This entire process --- i.e., the determination of routes and next hops, the passing down to lower-level scopes, and the passing up to intermediary nodes --- can be repeated as many times as required. In this example, the same process continues by passing the Bundle down once more, now to Scope 3, where the next level of routing takes place. This effectively models the recursive character and the scope-based scalability of the architecture presented herein.

\subsection{Relationship to the Region Concept}\label{concept-regions}

It shall be noted that another approach to network separation is currently being discussed in the DTN domain: the \emph{region} concept \cite{burleigh-regions-00}. The approach investigated in the REDMARS2 project is distinct, as it performs a \emph{vertical} subdivision of networks (into distinct scopes on different layers, which do not share any topological information) --- whereas the region concept can be considered a \emph{horizontal} approach (operating on the same layer, interconnecting regions using \emph{passageways} that have topological information from both regions available).
Both approaches can be considered complementary and may be combined without limitations. It is up to future work to determine which approach best fits a given set of use cases.

\section{Implementation}
\label{implementation}

\subsection{Recursive Networking Stack Based on BIBE} \label{ud3tn-bibe}
As outlined in the preceding sections, RINA relies heavily on scope-based rather than function-based separation. To clearly define these scopes and ensure that routing information is disseminated only within its relevant scope, a convergence layer protocol called Bundle-in-Bundle-Encapsulation (BIBE) is employed. This BIBE mechanism is implemented in µD3TN \cite{ud3tn2024} as a conventional Convergence Layer Adapter (CLA). The processing of BIBE Bundles is handled by two µD3TN instances, one representing the lower and one the upper Bundle Layer. In the RINA context, these Bundle Layers reside in distinct scopes and have their own contacts and links. In architectures not based on RINA, two Bundle Layers are not necessarily treated as separate entities.

The process of Bundle forwarding operates as follows: whenever the lower layer receives a Bundle containing a BIBE Protocol Data Unit (BPDU), it forwards the BPDU to the BIBE-CLA of the upper Bundle Layer via its application interface (AAP in case of µD3TN). The upper layer subsequently parses the Bundle encapsulated within the BPDU and processes it further. Given the objective of constructing a RINA-like architecture, the upper Bundle Layer constitutes an independent BPA instance within a scope distinct from the lower layer. This convention applies throughout the remainder of this document.

Once the Bundle is received by the upper layer, further processing depends on its destination. Two potential outcomes exist:
\begin{itemize}
    \item The destination of the encapsulated Bundle is an application registered on the upper layer.
\end{itemize}
or
\begin{itemize}
    \item The destination of the encapsulated Bundle is an application registered on a different node.
\end{itemize}
If the BPA determines that the Bundle's destination is a locally registered application on the upper layer, it simply forwards the Bundle to that application, and no further BIBE-related processing occurs. It should be noted that this application could potentially be the BIBE-CLA of a second upper-layer µD3TN instance; in this scenario, the process described here would be repeated following the forwarding of the BIBE Bundle.

\begin{figure}[h]
    \centering
    \includegraphics[width=0.5\textwidth]{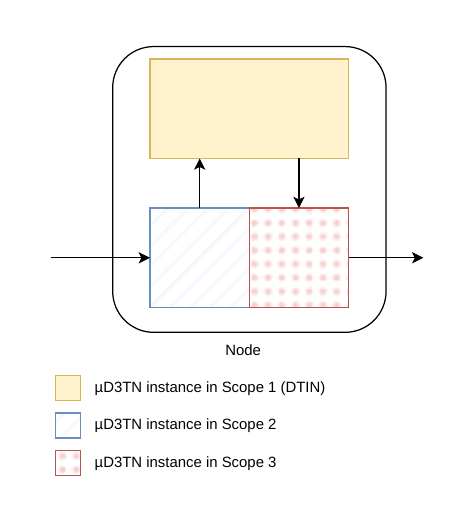}
    \caption{Illustration of BIBE Bundle forwarding via a node that is a member of multiple scopes. Note that the Bundle is not returned to the same µD3TN instance that forwarded the Bundle to Scope 1.}
    \label{fig:rina-with-bibe}
\end{figure}

However, if the BPA determines that this node is not the Bundle's final destination, its next hop is determined according to the local routing policy and the Bundle is forwarded to the next hop via the CLA selected for it. If the forwarding action is performed via the BIBE-CLA, the Bundle is re-encapsulated and sent back to the underlying scope using the application interface of the lower-layer µD3TN instance. Importantly, this interface is not necessarily the same from which the previous encapsulating Bundle originated (cf. Figure~\ref{fig:rina-with-bibe}). Moreover, this forwarding mechanism always constitutes a change of scope. Upon receiving the BIBE payload (the serialized upper-layer Bundle), the lower layer's Application Agent constructs a new Bundle, encapsulating it. The lower-layer Bundle processing routines then determine the next hop and the CLA to be used for onward transmission, after which the Bundle is forwarded accordingly.\par

\subsection{Distribution of Topology Information} \label{ud3tn-topological-info}
In the case of µD3TN, the distribution of topology information is dependent on the Bundle Dispatcher Module (BDM) implementation. In our prototypical implementation, which we used for evaluation in the REDMARS2 project, information concerning which destination node is reachable via which next hop is configured statically. In this specific case, crossing scope boundaries is not possible because topology information is not dynamically exchanged. The only point at which topology information can transition between scopes is during a Bundle transfer between them. However, since this transfer occurs via the BIBE-CLA, there is no risk of accidentally leaking topology information into the wrong scope in this process.

\section{Evaluation}\label{evaluation}
\subsection{Evaluation Scenario and Goals} \label{eval-scenario}
\begin{figure}[H]
    \centering
    \includegraphics[width=\textwidth]{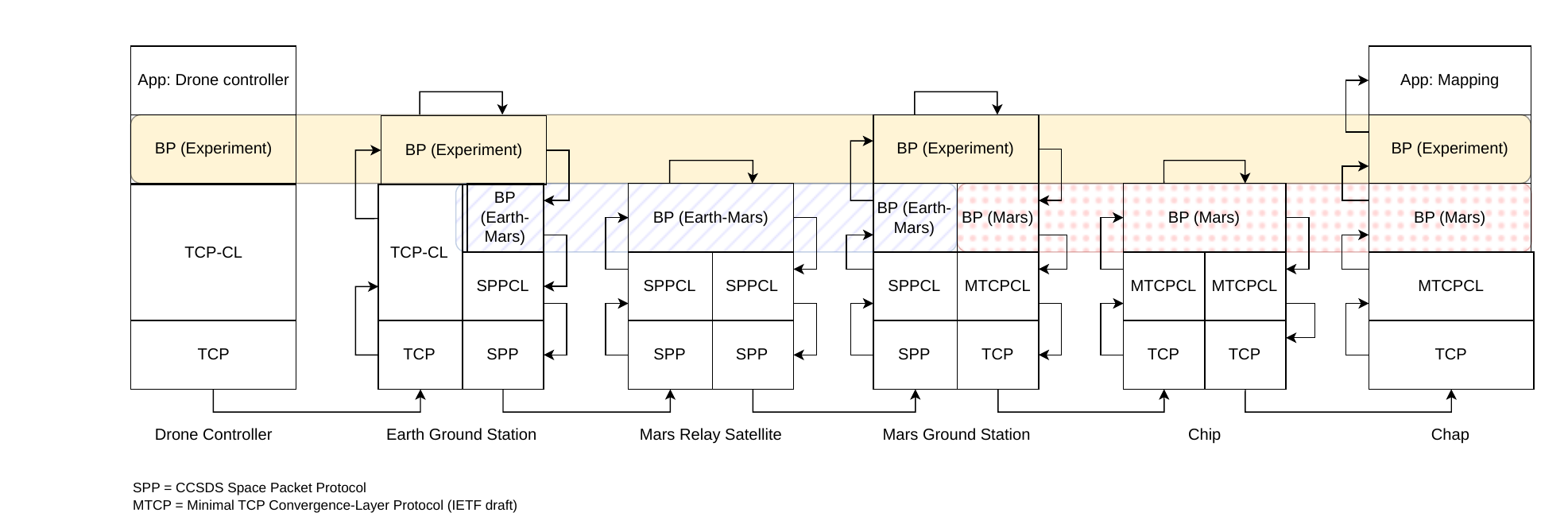}
    \caption{Schematic representation of the individual nodes in the evaluation scenario and the path that Bundles take from the source to the destination node.}
    \label{fig:scenario}
\end{figure}

To evaluate the implementation, a scenario was developed that models as many characteristics of RINA-enabled networks as possible. The final scenario is sketched in Figure~\ref{fig:scenario}. The underlying theme of the scenario is the communication between an operator on Earth and UAVs on the Martian surface. This scenario involves six nodes distributed across three scopes, which are briefly described below.

\paragraph{Experiment} The End-to-End Overlay Scope encompasses four of the six nodes. Alongside the drone operator’s computer and the ground station on Earth, the ground station and the Survey UAV on Mars are also part of this scope. Since this scope possesses the highest level of abstraction, routing for the Earth-to-Mars and Mars-to-Mars communication must accordingly defer to the other two scopes. 
\paragraph{Earth-Mars} Serving as the scope that connects the ground station on Earth with the ground station on Mars, this scope stores information that enables communication between the aforementioned ground stations. As the operational range of this scope is strictly limited to Earth-Mars communication, it is not possible to address destinations on Mars directly within this scope. 
\paragraph{Mars} The Mars Scope is responsible for planetary routing on Mars. Therefore, the only participants in this scope are the Mars ground station and the two UAVs stationed on Mars.  \\

The fundamental objective of the scenario is to transmit commands from Earth to a drone that, based on these commands, flies to the provided coordinates, captures a photograph, and sends it back to Earth along the same path. 
\subsection{Evaluation Setup} \label{eval-setup}
Due to the lack of actual Martian assets, the evaluation scenario was emulated on Earth. The setup included an operator PC alongside a Raspberry Pi 3B, which served as the Mars Ground Station. Additionally, each UAV was equipped with a Raspberry Pi 3B running the µD3TN protocol implementation, simulating the Mars reconnaissance drones. While it was initially planned to include a space component using the ESA OPS-SAT, delays for several reasons (including alignment of the conducted field tests with experiment availability periods, satellite passes, and weather conditions at the test site) prevented its use. Hence, we tested only with the OPS-SAT Engineering Model (FlatSat) on Earth.

\begin{figure}[H]
    \centering
    \includegraphics[width=\textwidth]{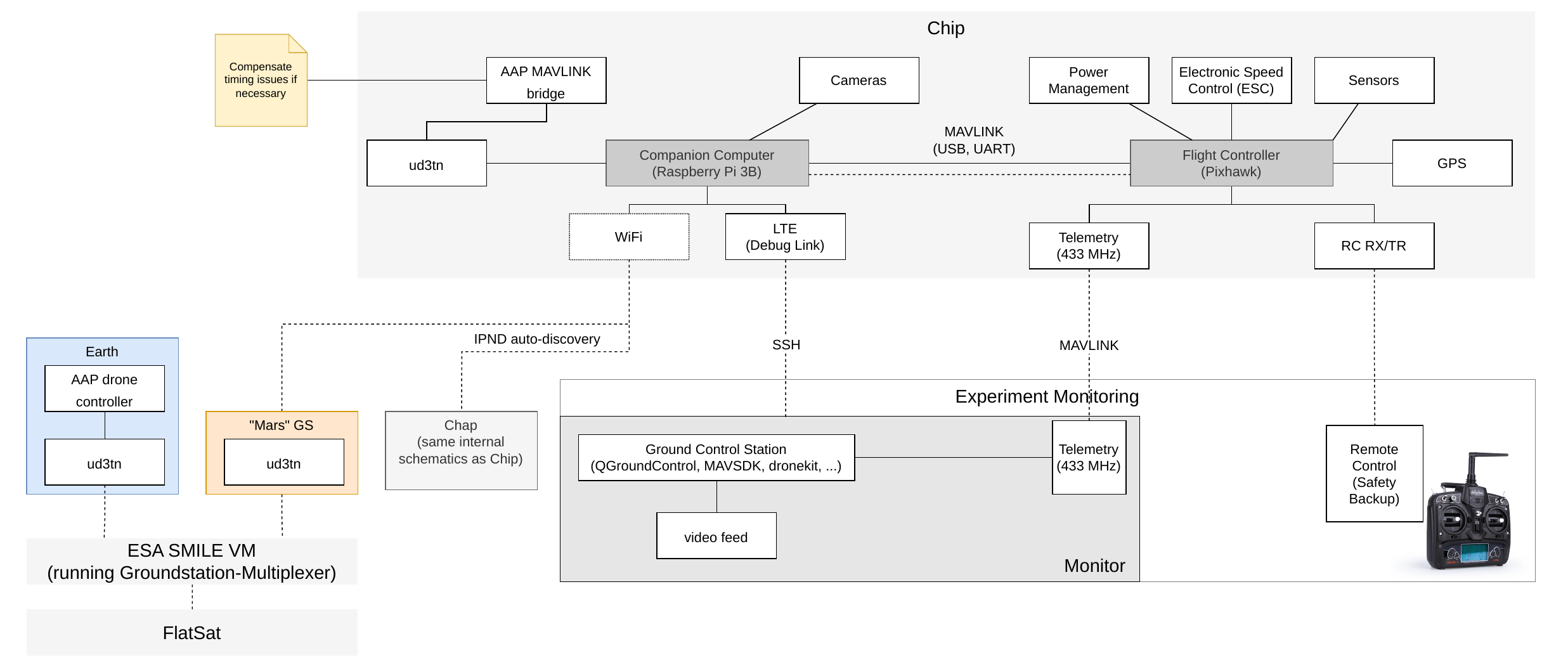}
    \caption{Visualization of the experimental setup, including the services and functionalities localized on the drones, as well as the monitoring. The remote controllers were only connected to the drones for safety backup and were not used for control during the field test.}
    \label{fig:scenario_setup}
\end{figure}

The exact experimental setup is illustrated in Figure~\ref{fig:scenario_setup}. A command is sent from the operator PC to the virtual machine (VM) running our ground station multiplexer. This multiplexer allows us to emulate multiple ground stations, using only a single physical station. From this VM, the command is forwarded to a satellite (in this case, the FlatSat), which subsequently sends it back. In the meantime, the ground station multiplexer has changed its configuration to emulate a different ground station. The command is now forwarded to the Raspberry Pi designated as the Mars ground station instead of back to the operator PC. This emulates the transmission of the command from Earth to Mars. In the REDMARS2 scenario, the command is then transmitted to the drone \say{Chip} located near the Mars ground station, which independently determines the last known location of the target node (the drone \say{Chap}) and flies toward it. Using DTN IP Neighbor Discovery (IPND), a contact is established between the two UAVs, enabling the operator's command to be transmitted. Chap then evaluates this command and flies to the specified coordinates to capture a photograph. Once the task is complete, the photograph is transmitted back to the operator via the same path.

The integration of RINA concepts into this experimental setup is primarily based on BIBE, as previously described. This means that the scopes highlighted by color in Figure~\ref{fig:scenario} do not have to be explicitly defined. Instead, they are implicitly formed by configuring the contacts and convergence layer links using the BIBE-CLA and providing the corresponding routing and topology information to the individual BPA instances.

\begin{figure}[H]
    \centering
    \includegraphics[width=\textwidth]{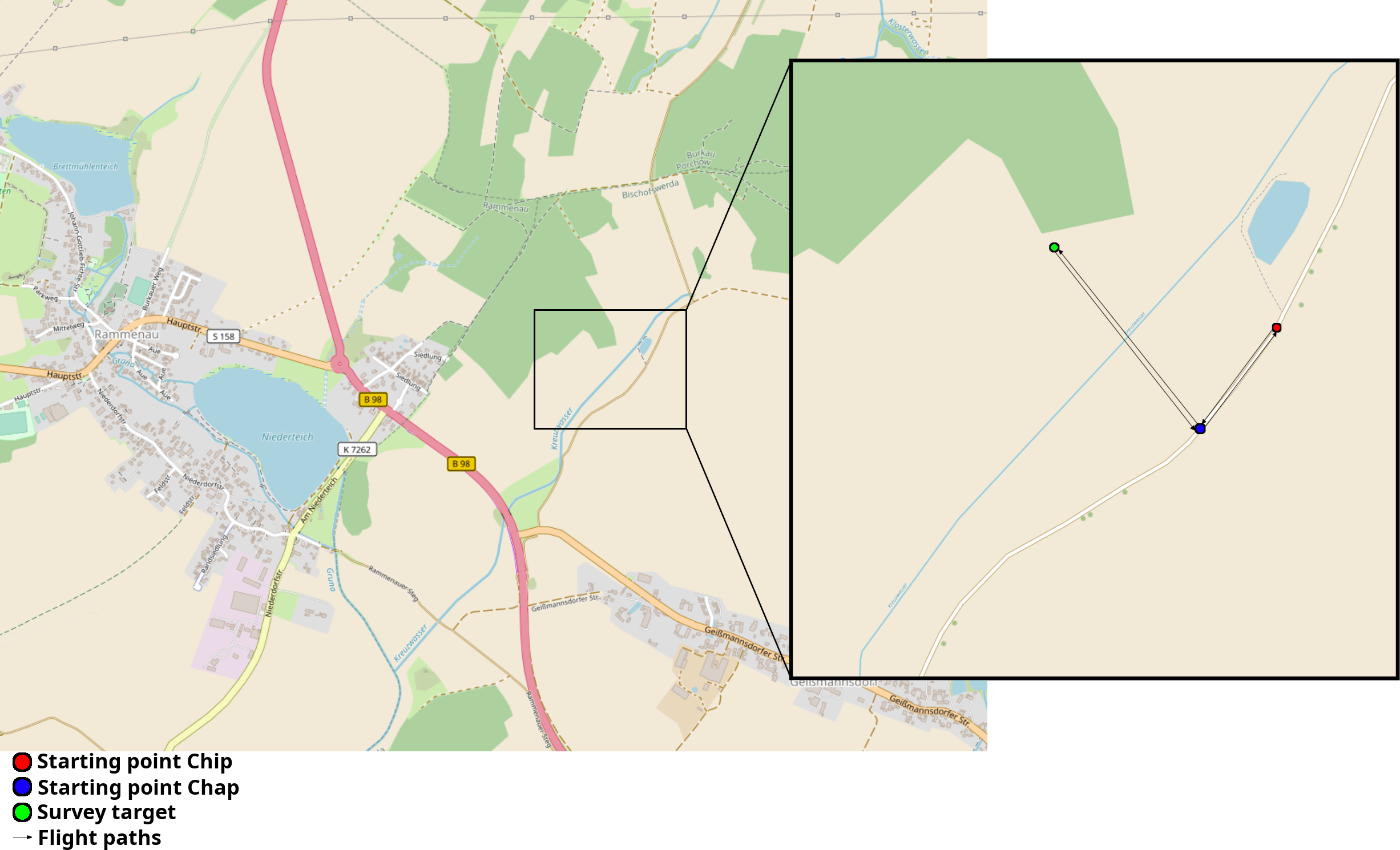}
    \caption{Starting points, flight routes, and target of the UAVs during the field test on 2022-09-22}
    \label{fig:map}
\end{figure}

\begin{figure}
\begin{tabular}{p{76.2mm}p{76.2mm}}
  \includegraphics[width=76.2mm]{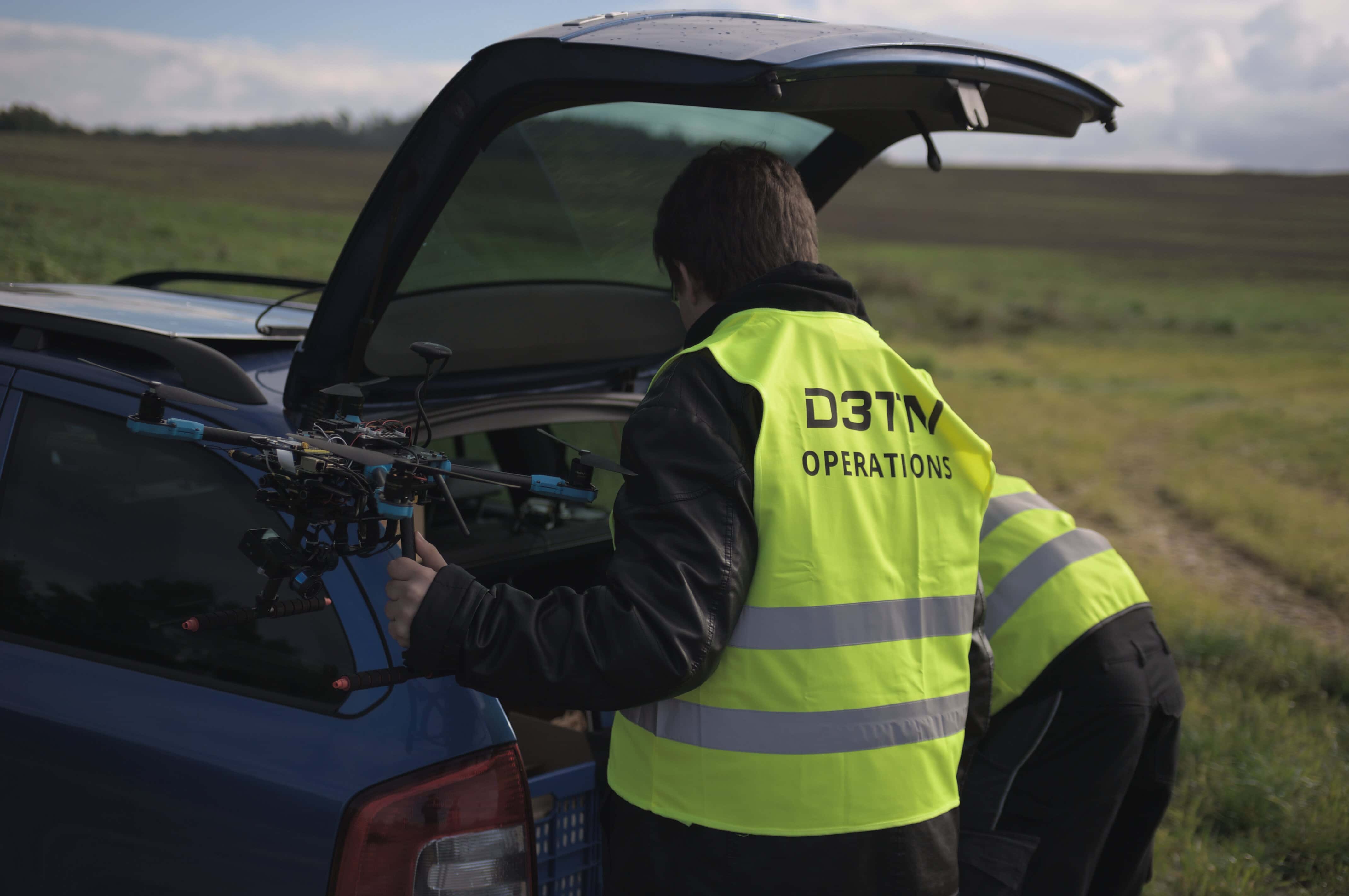} &   \includegraphics[width=76.2mm]{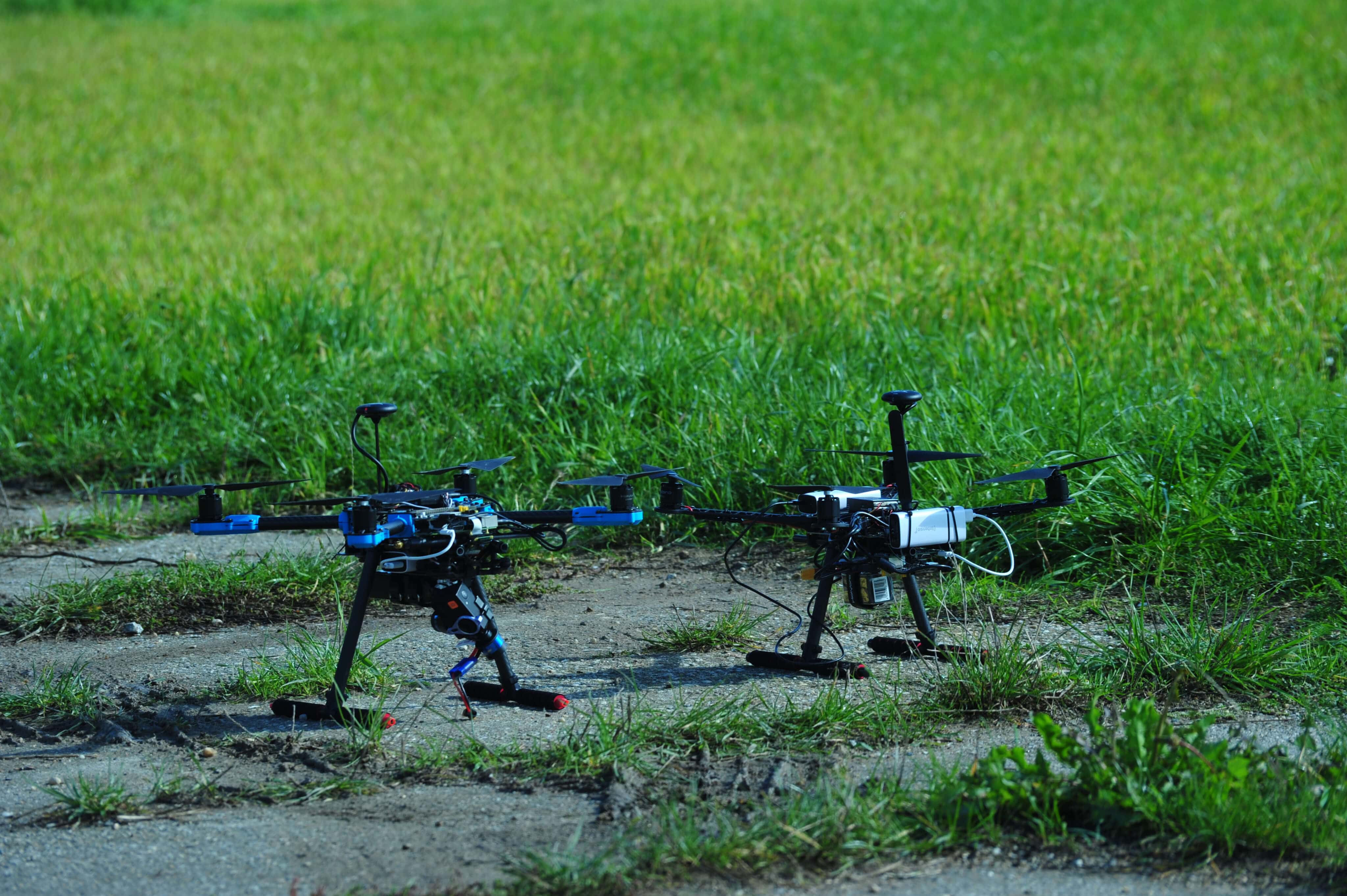} \\
(a) Setup of the experimental configuration & (b) The two UAVs \say{Chip} (left) and \say{Chap} (right) \\[6pt]
 \includegraphics[width=76.2mm]{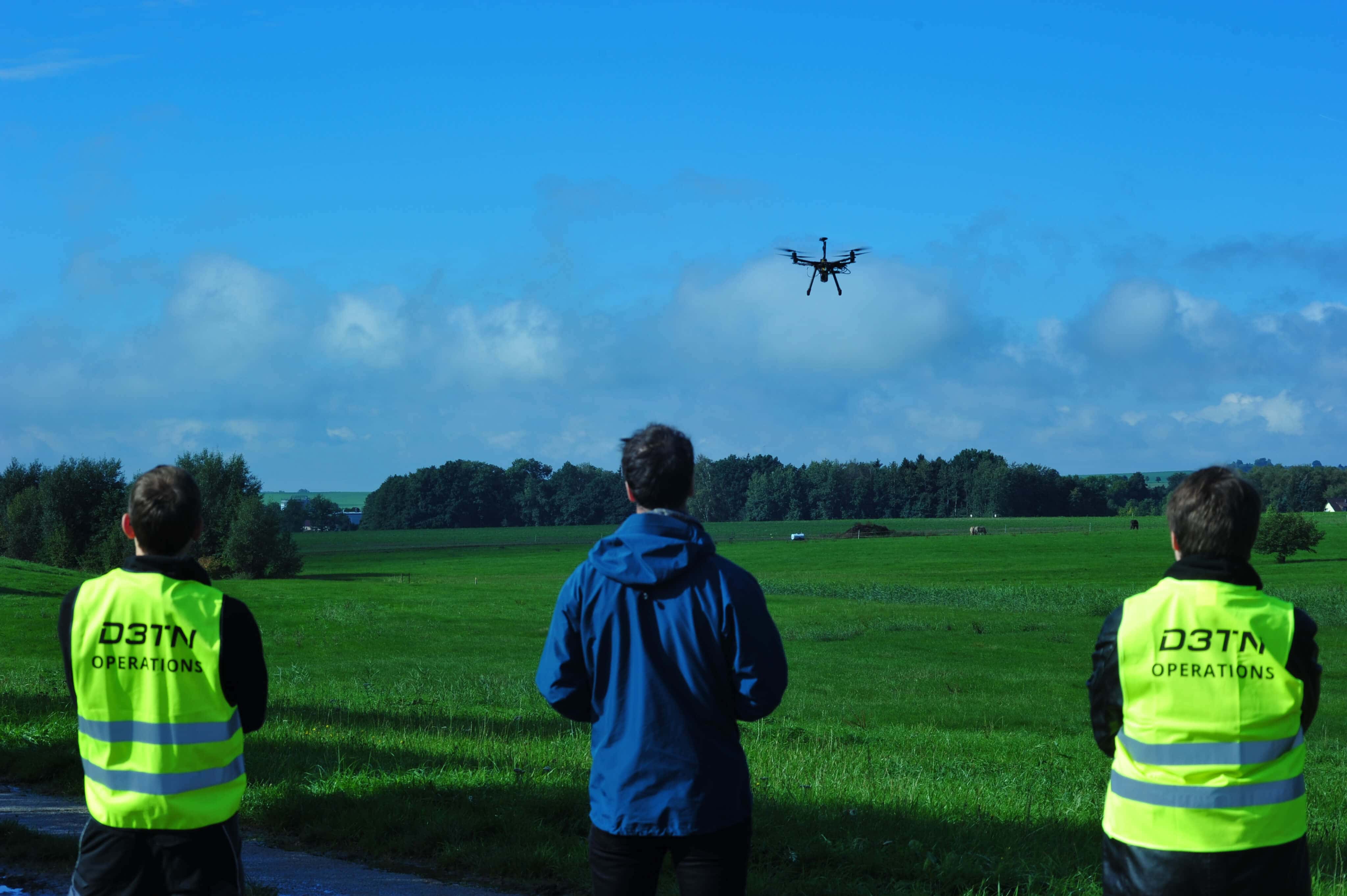} &   \includegraphics[width=76.2mm]{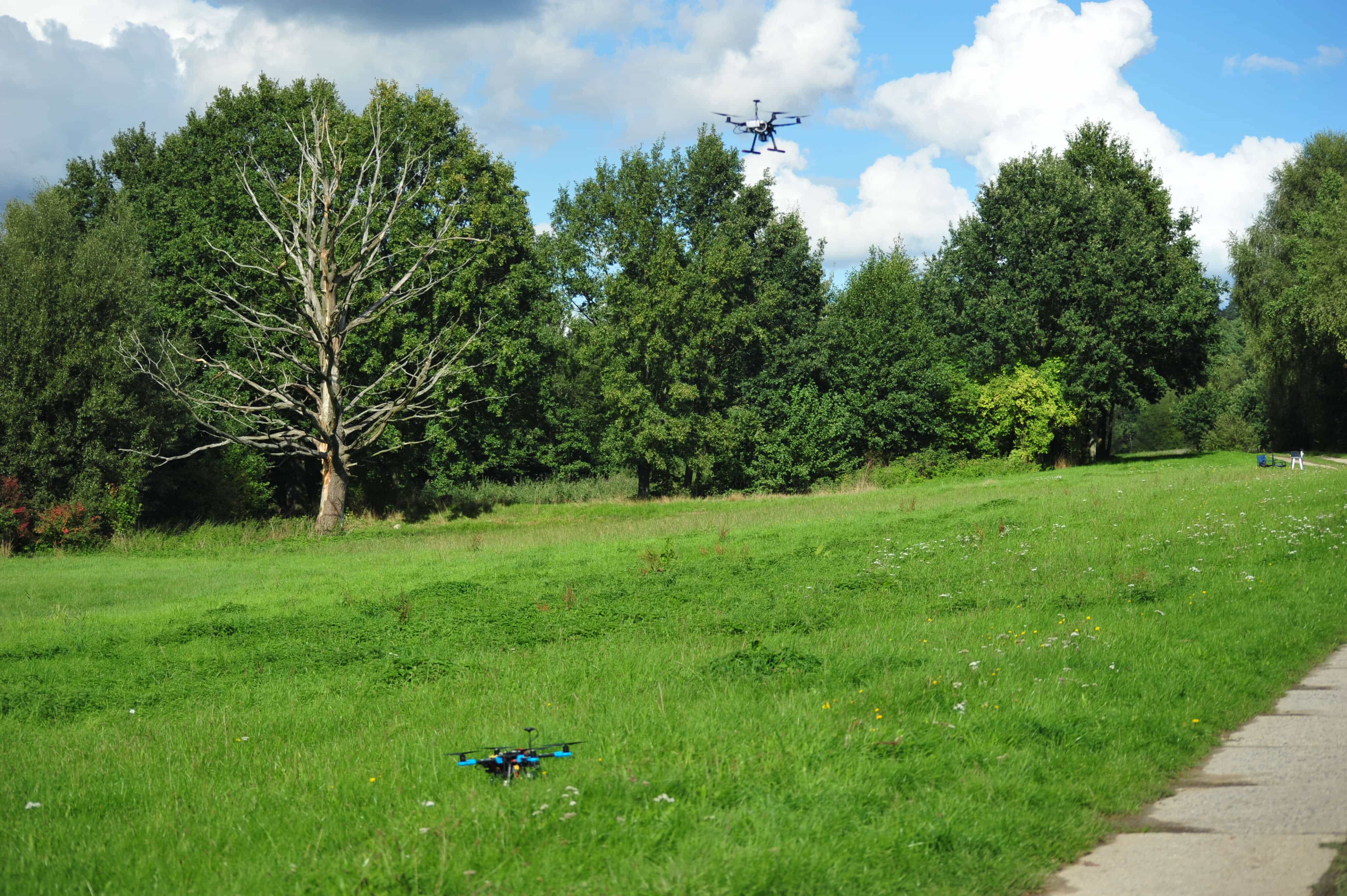} \\
(c) Chip on its way to Chap & (d) After the mission data transfer from Chip (on the ground) to Chap, Chap begins mission execution \\[6pt]
\includegraphics[width=76.2mm]{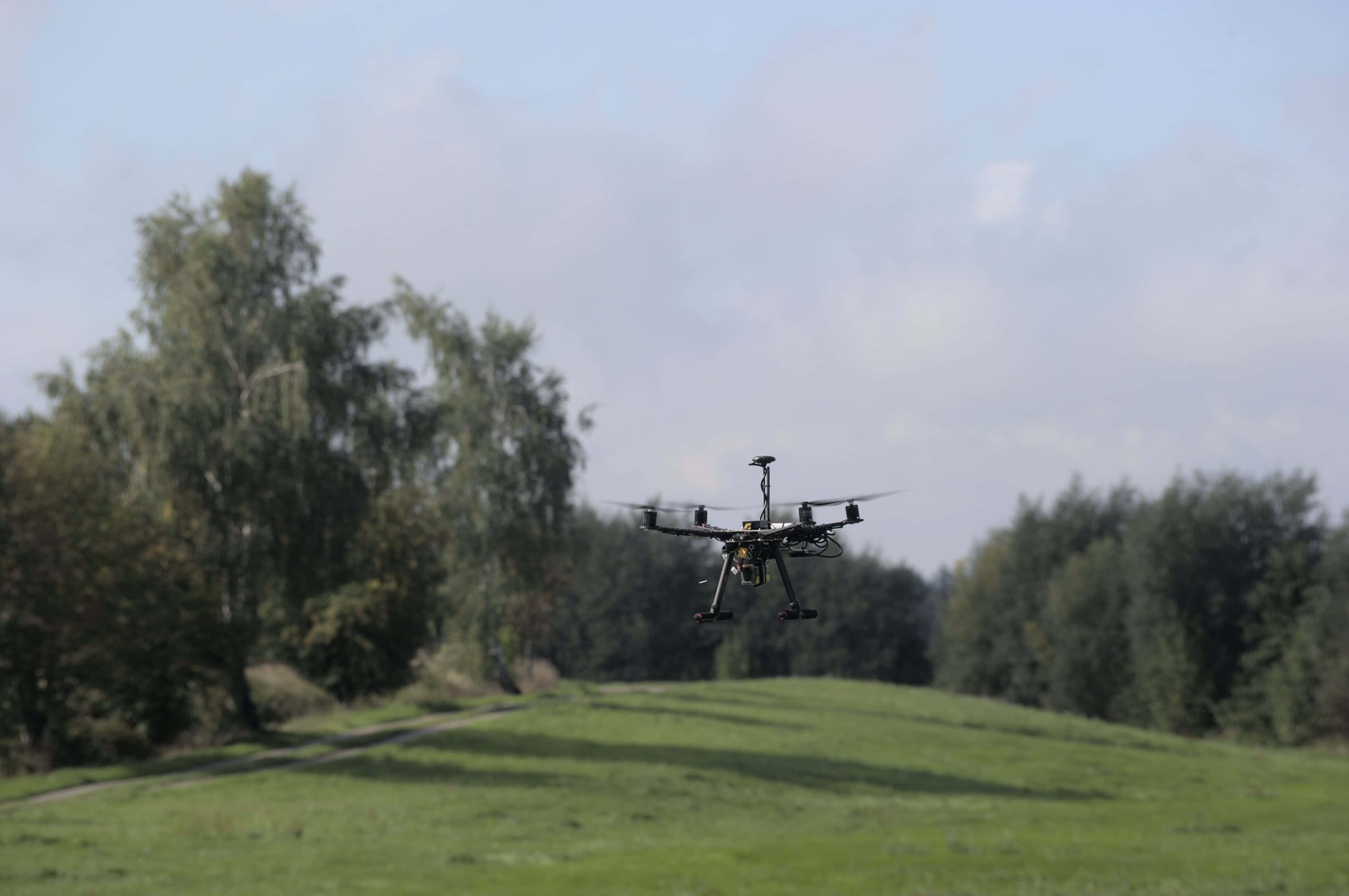} & \includegraphics[width=76.2mm]{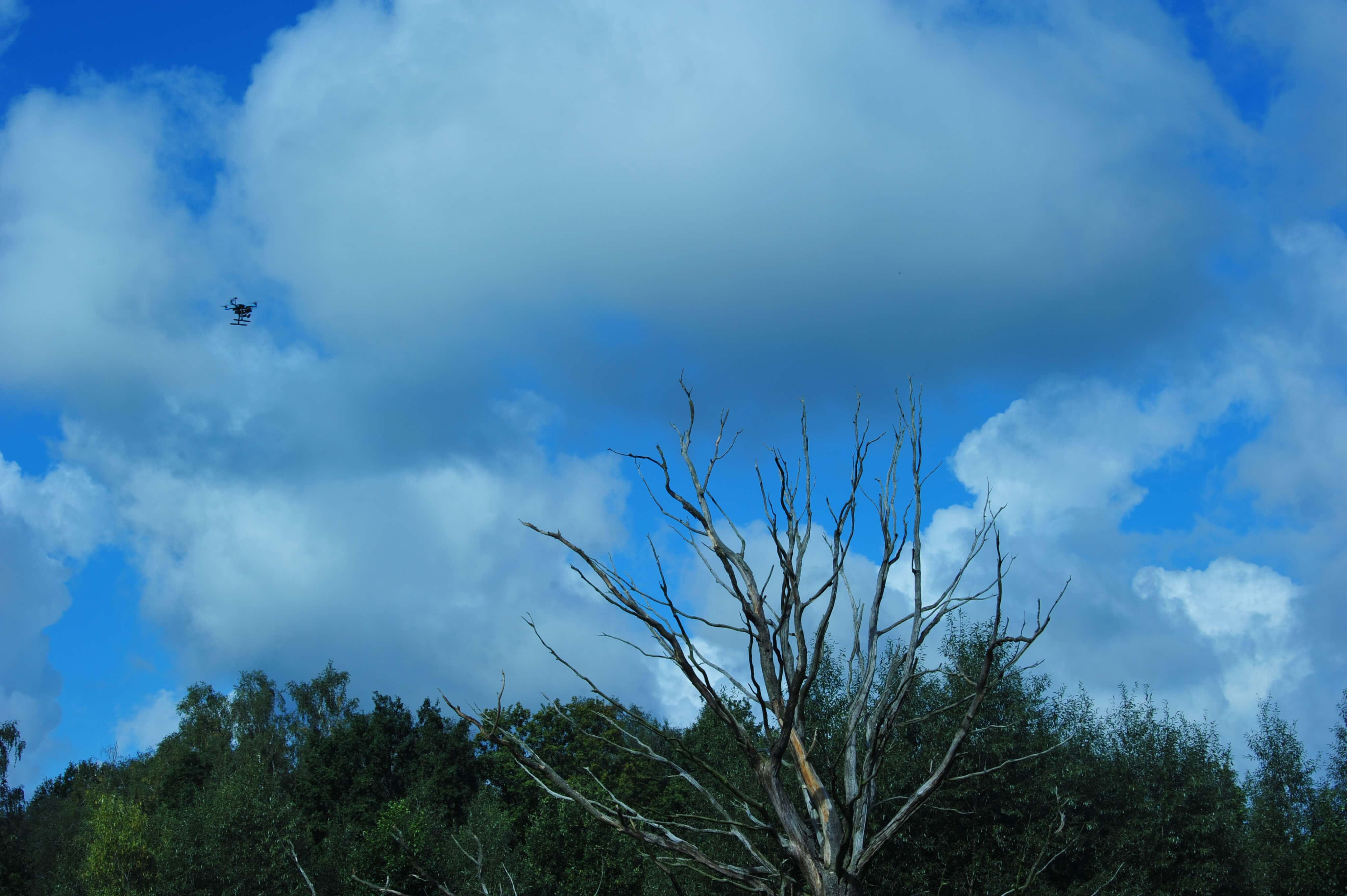}\\
(e) Chap flying to the target coordinates & (f) Chap capturing a photograph of the target \\

\end{tabular}
\caption{Procedure of a test run during the field test conducted on September 22, 2022}
\label{fig:field_test}
\end{figure}

The described experimental setup was tested and evaluated during several field tests, both with and without ESA asset integration. The main test, which verified the seamless interaction of all functions in conjunction with the ESA FlatSat, took place on September 22, 2022. Figure~\ref{fig:map} provides an overview of the test environment, including the starting positions, flight paths, and target objects of the drones. Furthermore, Figures~\ref{fig:field_test}a–f illustrate the successful execution of the experiment during the main test day.

\subsection{Evaluation Results} \label{eval-results}

During the field tests described in the previous section, we were able to follow the procedure outlined in the evaluation setup. The transmission of commands to the UAV \textit{Chap}, its execution of the task, and the return of the recorded photo via the ESA FlatSat all proceeded smoothly. No routing or topology information was found to have been transferred to scopes for which it was not intended. 

All in all, the usage of RINA concepts within DTN can be considered successful without limitations. Among other findings, we showed that the relevant concepts can be transferred to DTN without significant issues. Specifically, the usage of BIBE rendered it possible to bridge heterogeneous DTN underlay networks via an DTN overlay network providing end-to-end Bundle transmission capabilities. The transferred concepts could enable a highly scalable DTN with minimal effort, with applications in domains such as the Solar System Internet.

\section{Summary and Outlook}\label{summary}

This whitepaper has summarized key results from the REDMARS2 project conducted in 2021 and 2022. The presented findings focus on the application of BIBE for forming scalable DTNs. The core idea is to introduce separate topological scopes by operating DTN overlay networks on top of one or more DTN underlay networks. As discussed in this paper, BIBE constitutes the fundamental technology that enables this type of layering. In addition to the conceptual discussion, a reference implementation leveraging µD3TN was presented. The proposed concepts were validated in a field test demonstrating a remote drone control system applicable via heterogeneous DTNs.

Future work may focus on analyzing the scalability of the proposed approach in more detail.

\newpage

\printbibliography
\end{document}